\documentclass[12pt,dvips]{article}
\usepackage{color}
\usepackage{graphicx}
\usepackage{latexsym}
\usepackage{makeidx}
\usepackage{dafne99pro}

\newcommand{\gev}{\, {\rm GeV}}

\newcommand{\bea}{\begin{eqnarray}}
\newcommand{\eea}{\end{eqnarray}}
\newcommand{\be}{\begin{equation}}
\newcommand{\ee}{\end{equation}}
\newcommand{\bi}{\begin{itemize}}
\newcommand{\ei}{\end{itemize}}

\newcommand{\beq}{\begin{equation}}
\newcommand{\eeq}{\end{equation}}
\newcommand{\ba}{\begin{array}}
\newcommand{\ea}{\end{array}} 
\newcommand{\beqa}{\begin{eqnarray}}
\newcommand{\eeqa}{\end{eqnarray}}

\newcommand{\gsim}{\stackrel{>}{_\sim}}

\newcommand{\hepph}[1]{\texttt{hep-ph/#1}}
\newcommand{\hepex}[1]{\texttt{hep-ex/#1}}

\begin{document}

\title{Overview of Chiral Perturbation Theory}

\author{
Gilberto Colangelo \\
{\em Institut f\"ur Theoretische Physik der Universit\"at Z\"urich} \\
{\em Winterthurerstrasse 190, CH--8057 Z\"urich}}

\maketitle

\baselineskip=11.6pt
\begin{abstract}
I present a quick overview of the current status of Chiral perturbation
theory in the meson sector. To illustrate the successes and some problems
in the description of the phenomenology, I focus on a few selected examples
that are relevant for DA$\Phi$NE.
\end{abstract}
\baselineskip=14pt
\section{Introduction}
Chiral perturbation theory (CHPT) is the low--energy effective theory of
the strong interactions. To characterise quantitatively the meaning of
``low energy'' in this framework, we recall that the relevant physical
scale here is that of spontaneous chiral symmetry breaking, i.e. about 1
GeV: the effective theory is supposed to work only 
for $E \ll 1$ GeV. For practical purposes one expects CHPT to work
reasonably well up to 500--600 MeV. Which means that at DA$\Phi$NE, as soon
as the $\phi$ decays we enter the realm of CHPT: for any physical process
occurring after the decay of the $\phi$ there is most likely some relevant
piece of information that can be derived from CHPT (a check on this claim
can be easily made by glancing through the ``Second DA$\Phi$NE Physics
Handbook''\cite{handbook}).

This effective field theory is a systematic extension of the
current--algebra methods that were used in the sixties in hadronic
physics. Its present form is due to Weinberg\cite{weinberg79} and Gasser
and Leutwyler\cite{GL}. They showed the advantages of the
effective--field--theory language over the direct implementation of the
Ward identities as in the current--algebra framework. In particular because
what was known before as a tremendously difficult problem, the calculation
of the corrections to a current--algebra result, was reduced to a routine
loop calculation in a well--defined framework.
After the convenient tools of the effective field theory were made
available, many processes have been calculated at the one--loop level:
wherever possible, the comparison to the experimental data has shown a
remarkable success of the method.

In the early nineties, also because of the prospects of having a
$K$--factory operating soon, the first two--loop calculations were
made. The first one was the cross section for the two--photon annihilation
into two neutral pions\cite{BGS}.  This beautiful and difficult calculation
opened up the field of two--loop calculations in CHPT. In fact if we
consider only the two--light--flavour sector, all the phenomenologically
relevant calculations have already been done, whereas in the $SU(3)$
framework, they are just starting\cite{ABT}. Moreover, in the purely strong
sector, the Lagrangian at order $p^6$ and the complete divergence structure
have been recently calculated\cite{lagrp6}. What I find remarkable is that,
despite the rapidly increasing number of new constants appearing at each
new order\cite{GL,lagrp6}, the theory is able to produce sharp predictions,
like in the first instance, the calculation of the $\gamma \gamma \to \pi^0
\pi^0$ cross section. The best illustration of this is the $\pi \pi$
scattering reaction, that I will discuss in more details in the following
section.

In parallel to these two--loop calculations arose the need to account for
small effects such as electromagnetic corrections and purely strong
isospin--breaking effects. While the latter are readily calculated with the
Lagrangian of Gasser and Leutwyler, the former need the inclusion of the
electromagnetic field in the theory. Loops with a virtual photon field
generate new types of divergences, that need new counterterms to be
removed. Such a Lagrangian was formulated by Urech, and Neufeld and
Rupertsberger\cite{urech} at order $p^4$ for the strong sector.  For
processes involving also leptons one needs a further extension of the
Lagrangian to account for contributions of virtual leptons inside the
loops. This has been formulated only very recently\cite{knecht}, opening up
the way to phenomenological applications, such as semileptonic kaon
decays. This is an essential step forward if we want to fully exploit the
precision of the data on $K$ decays that experimentalists are starting to
provide\cite{lowe,kettell}.

The application of CHPT to weak nonleptonic decays is more problematic
because of the presence of more constants already at order $p^4$. In
addition, there are less measured quantities from which to extract these
constants. This means, e.g., that in the classical sector of the kaon
decays into two or three pions CHPT has rather little to say (recently
there has been a very nice counterexample\cite{omega} to this
statement). The situation improves if one looks at the
radiative--nonleptonic--decay sector, where the theory can make predictions
ands can be meaningfully tested. I will discuss this topic in
Sect. \ref{sec:weak}.

Conceptually, the basic ingredients in the formulation of CHPT are the
spontaneous (global) symmetry breaking and the existence of a mass gap in
the spectrum between the Goldstone modes and the other energy levels.
The effective field theory for such a situation can be formulated in very
general terms, without any reference to a specific symmetry group,
or a specific physical system. This method has in fact been applied to a
variety of different physical systems and situations, ranging from
solid--state systems to strong interactions at finite temperature and
volume, to QCD in the quenched approximation, etc.
I will not discuss these different fields of applications (I refer the
interested reader to the excellent reviews available in the literature
\cite{reviews}), and restrict myself to applications in the meson sector
only, leaving out also the very important and rich field of baryon physics. 

\section{Phenomenological applications: strong sector}

Phenomenological applications in the strong sector are nowadays at the
level of two--loop calculations. $\pi \pi$ scattering\cite{BCEGS} is the
best example to show what high level of precision one can aim to with
two--loop CHPT. The scattering lengths, predicted in CHPT, can be measured
in $K_{e4}$ decays, but also with the help of pionic atoms. It is known
since many years that the lifetime of pionic atoms is proportional to the
square of the difference of the two $S$--wave scattering lengths, modulo
corrections. A precise evaluation of these corrections is crucial if one
wants to pin down the scattering lengths at a few percent level. CHPT
can help also in this case. It is a beautiful example of the power of the
method, which also shows that the field is still open to progress at the
methodological level, not only via multiloop calculations.

\subsection{$\pi \pi$ scattering}
\begin{figure}[t]
\begin{center}
\leavevmode
\includegraphics[width=\textwidth]{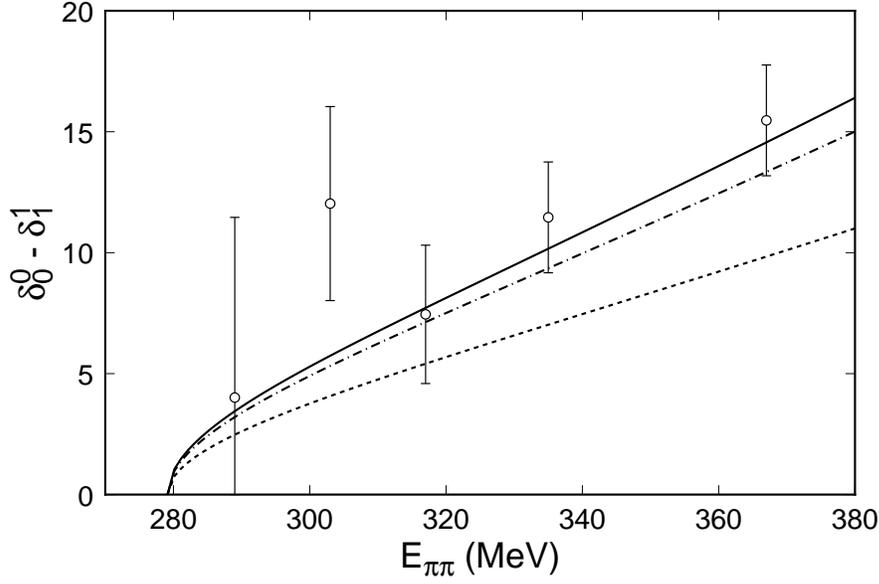}
\caption{\label{fig:ke4} {\it
Data: Geneva--Saclay collaboration\protect{\cite{rosselet}}
Dashed line: CHPT at $O(p^2)$ 
Dot--dashed line: $O(p^4)$
Solid line: $O(p^6)$} }
\end{center}
\end{figure}

This is the ``golden reaction'' for Chiral Perturbation Theory: at
threshold the naive expansion parameter is $M_\pi^2/1 \gev^2 \sim 0.02$,
and already a tree level calculation\cite{weinberg66} should be rather
accurate. This rule of thumb is quite misleading here, as it is shown by
the fact that both the one--loop\cite{GL83} and the two--loop\cite{BCEGS}
calculations produced substantial corrections. The violation of the rule of
thumb has a well known origin, and is due to the presence of chiral
logarithms $L= M_\pi^2/(4 \pi F_\pi)^2 \ln M^2/\mu^2$, which, for $\mu \sim
1$ GeV change the expansion parameter by a factor four. If we look at the
$I=0$ $S$--wave scattering lengths, e.g., a large coefficient in front of
the single (at one loop) and double (at two loops) chiral logarithms is the
main source of the large correction\cite{GL83,GC95}:
\begin{eqnarray}
a^0_0 &=& \frac{7 M_\pi^2}{32 \pi F_\pi^2} \left\{1 -
  \frac{9}{2}L  + \frac{857}{42} L^2 + \ldots \right\} \nonumber \\
&=& \underbrace{\overbrace{ 0.156}^{\mbox{tree}}
  +\overbrace{ 0.039 +
0.005}^{\mbox{1~loop}}}_{\mbox{0.201}}+
\overbrace{ 0.013+0.003+0.001}^{\mbox{2~loops}} \nonumber \\
&=& \underbrace{0.217}_{\mbox{total}}
\end{eqnarray}
The same picture is maintained if we move away from threshold. In
Fig. \ref{fig:ke4} one can see the comparison of the three successive
chiral orders and the experimental data for the phase--shift difference
$\delta_0^0-\delta_1^1$ coming from $K_{e4}$ decays.  The figure shows a
well behaved series that ends up in rather good agreement with the
experimental data. At this level of accuracy of the experimental data,
however, the comparison is not particularly instructive, and even a precise
assessment of the theoretical uncertainties would not seem necessary. The
real challenge comes from the present generation of experiments: both KLOE
at DA$\Phi$NE and E865 in Brookhaven\cite{lowe} will be able to
analyse a factor ten more events than the old Geneva--Saclay
collaboration\cite{rosselet}. Without taking into account the improvements
in the systematics (which should be particularly important for KLOE in view
of its very clean environment), the reduction of the error bars is of
about a factor three. Which makes it a real precision test.

To make the discussion of the numerics a little simpler it is useful to
come back to the scattering length. To compare theory and experiment here
we first have to solve the problem of the extraction of the scattering
length from the measurement of the phase shift. Can this be done reliably,
without introducing further uncontrolled uncertainties? The answer is
positive, and the method to do this relies on solving numerically the Roy
equations. The latter embody in a rigorous way the analyticity and
crossing--symmetry properties of the $\pi \pi$ scattering amplitude -- when
supplemented with the unitarity relations they become nonlinear, and
amenable only to numerical studies. The physical $\pi \pi$ scattering
amplitude must obviously satisfy them. These equations have two subtraction
constants: the two $S$--wave scattering lengths. If one specifies the
values of these two subtraction constants (and also uses experimental input
at high energy, $E> M_\rho$), the solution is unique. One may reverse the
argument and say that the physical amplitude away from threshold (as
measured experimentally in $K_{e4}$ decays) determines unambiguously the
two scattering lengths. 
\begin{figure}[t]
\begin{center}
\leavevmode
\includegraphics[width=10cm]{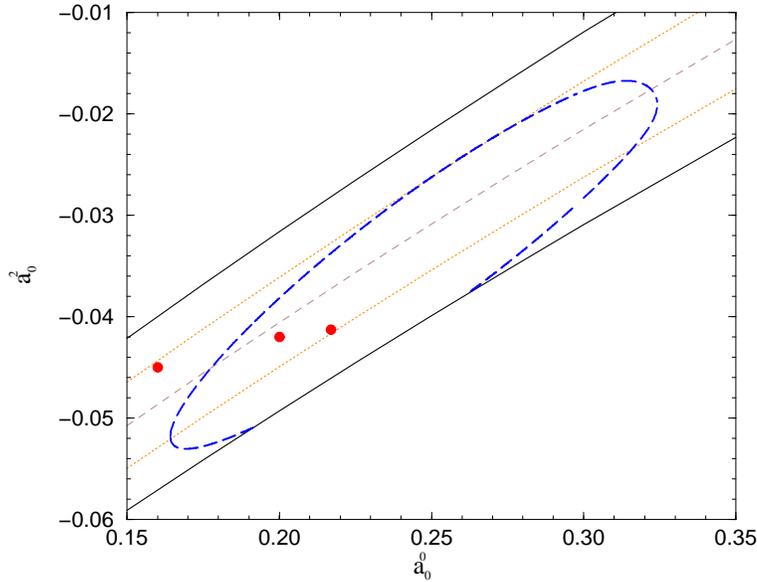}
\caption{\label{fig:ellipse}{\it The contour corresponds to a 70\% C.L., the
  three points indicate the three successive chiral orders. The band inside
  the two solid lines is the so--called Universal Band: no solutions of Roy
  equations may be found outside the band.}}
\end{center}
\end{figure}
Such a program had been carried out in the
seventies by Basdevant, Froggatt and Petersen\cite{BFP}. Today it has been
revived\cite{Roysolution} to be used again with the new generation of
experimental measurements.  Both the new\cite{Roysolution} and the old
analysis\cite{BFP} of Roy equations agree in that the data of the
Geneva--Saclay collaboration constrain the $I=0$ $S$--wave scattering
length to be roughly between $0.20$ and $0.30$ in pion mass units. This can
be seen in Fig. \ref{fig:ellipse}, where it is shown the 70\% C.L. contour
as obtained\cite{Roysolution} from the analysis of the available
experimental data in the low--energy region\cite{rosselet,hoogland}. The
chiral prediction is clearly well compatible, but in principle, a reduction
of the range to one third of its present size could show a discrepancy
between the experiment and the theory.  What would this mean?  Could the
theory change its numerical estimate by adjusting a few parameters here and
there?  A necessary ingredient to formulate an answer is a careful estimate
of the uncertainties to be attached to the two--loop prediction of
CHPT\cite{CGL}. This careful estimate, however, will say whether the final
uncertainty is 3, or 5 or 10\%, it is already very clear that there is no
way for CHPT to be in agreement with the central (or higher) value of
$a_0^0$ as preferred from the Geneva--Saclay data, $0.26$.

The only way to understand such a central value is a drastic change of
perspective at a very fundamental level: a value of the quark--antiquark
condensate much smaller than what we currently believe (and implicitly
assume in the formulation of CHPT) would increase substantially, already at
tree level, the value of $a_0^0$\cite{KMSF}:
\be
a_0^0 \gsim 0.26 ~~~ \Rightarrow ~~~ - {\langle \bar q q \rangle \over
  F_\pi^2} \ll 1 \gev \; \; .
\ee

\subsection{Decay of bound states}

An alternative method to measure experimentally the scattering lengths
uses pionic atoms, and therefore has the advantage of measuring the
interaction of pions right at threshold. The $\pi^+ \pi^-$ atom is an
electromagnetic bound state which decays predominantly into 
$2 \pi^0$ via strong interactions. The leading--order expression for the
width reads\cite{Deser}:
\be
\Gamma_{2 \pi^0}={2 \over 9} \alpha^3 p^* (a_0-a_2)^2, \; \; \; \; p^*=
\sqrt{M_{\pi^+}^2-M_{\pi^0}^2- \alpha^2/4 M_{\pi^+}^2}
\label{eq:width}
\ee
The DIRAC experiment at CERN aims to accurately measure the lifetime of
pionic atoms. The goal is a 10\% accuracy on the width, i.e. 5\% on the
scattering lengths. This simple and direct translation of the errors from
what is actually measured (the lifetime) to the scattering lengths
illustrates the tremendous advantage of the use of pionic atoms.
The advantage, however, would be a fake one, if the the leading order
expression for the lifetime (\ref{eq:width}) was subject to large and/or
difficult--to--control corrections.

There have been various attempts to calculate the corrections to the
formula (\ref{eq:width}), with several different methods\cite{Hadatom99}.
The results were, unfortunately, not always in mutual agreement, and due to
basic differences in the approaches, it seemed difficult to trace the origin
of the discrepancies. The situation has now changed, thanks to the
calculation of these corrections that was made in the framework of CHPT
\cite{bern_atom}. The calculation required the combined use of the CHPT
Lagrangian and the nonrelativistic Lagrangian method proposed by Caswell
and Lepage\cite{CL}. The result can be expressed in the following form:
\be
\Gamma_{2 \pi^0}={2 \over 9} \alpha^3 p^* {\cal A}^2 (1+ K)
\label{eq:chiral_width}
\ee 
where 
\be 
{\cal A} = a_0-a_2+h_1(m_d-m_u)^2+h_2 \alpha + o(\delta) 
\ee
\be 
K = \frac{M_{\pi^+}^2-M_{\pi^0}^2}{9M_{\pi^+}^2} (a_0+2 a_2)^2 - {2
\alpha \over 3} (\ln \alpha -1) (2 a_0+a_2)+o(\delta) 
\ee 
where $(m_u-m_d)^2$ and $\alpha$ are counted as small quantities of order
$\delta$. In the formulae above, $a_{0,2}$ are meant in the isospin limit
and for $M_\pi=M_{\pi^+}$.  The advantage of the use of CHPT is very clear:
the method does not only provide a number for the correction to the leading
order formula (\ref{eq:width}), but rather an algebraic expression, a
Taylor series expansion in $\delta$, with coefficients that can be calculated
unambiguously. If anybody wants to calculate these corrections with a
different method, he should also be able to compare the results for the
coefficients, and easily understand the origin of possible differences. One
could have in principle a more clever way to calculate these corrections,
summing up series of terms to all orders, but possibly not having the
complete leading correction: a comparison at the algebraic level would
easily clarify all these aspects. This program of detailed comparisons
between different methods has in fact already been started.  A handy
summary of the present status can be found in the MiniProceedings of
HadAtom99\cite{Hadatom99}, where the interested reader will also find
reference to the relevant literature.

\section{Phenomenological applications: weak sector}
\label{sec:weak}
In the weak--interaction sector the lowest order Lagrangian ${\cal
  L}_W^{(2)}$ contains only two constants: $c_2$ and $c_5$. The situation
is therefore similar to the one in the strong sector at this level.
At next--to--leading order the Lagrangian ${\cal L}_W^{(4)}$ contains
37 new constants called the $N_i$\cite{weak_lagr}.
Such a large number of constants seems to make the situation hopeless.
In the sector of $K$ decays into two or three pions, one can say that to a
large extent it is so. In the sense that it is difficult to make
predictions: one does not have enough observables from which to determine
some of the relevant constants, at best one can only fit the
data. Moreover, the simple method of resonance saturation to estimate the
values of the constants, does not seem to work in the weak sector as it
does in the strong one. At least not as straightforwardly.

The situation improves if one considers the nonleptonic radiative decays:
here only a restricted set of the constants contribute and there are many
more observables from which to determine them. In fact KLOE and the fixed
target experiments at Brookhaven and Fermilab are going to collect an
impressive number of data on many of these decays\cite{lowe,kettell}, and
will allow to determine rather precisely several of the $N_i$ constants. I
will concentrate in what follows on a couple of such decays, discussing the
importance of $O(p^6)$ contributions, when the order $p^4$ fails to
describe the data at the present level of accuracy. While a complete
formulation of the theory (knowledge of the complete Lagrangian and
divergence structure) at order $p^6$ in the weak sector is completely out
of sight at the moment, it is possible to push the calculation at the
$O(p^6)$ level, picking up only the presumably dominant terms.

\subsection{$K \to \pi \gamma \gamma$}

Assuming $CP$ conservation the $A(K_L\to \pi^0 \gamma \gamma)$ is
determined by two invariant amplitudes, $A(s,\nu)$ and $B(s,\nu)$,
$s=(q_1+q_2)^2, \nu=p_K\cdot(q_1-q_2)$, where $q_{1,2}$ are the momenta of
the two photons, and $p_K$ that of the kaon.  At order $p^2$: $A=B=0$.  At
order $p^4$\cite{Kpigg_p4}: $A = 4/ s (s-M_\pi^2) F(s / M_\pi^2) + \ldots$,
and $B=0$, where $F(x)$ is a loop function generated by $\pi \pi$
intermediate state in the $s$ channel, that represents the dominant effect
at this order, and the ellipsis stands for other less important
contributions. Although the shape of the spectrum was nicely confirmed by
the experiment\cite{E731,NA31}, the branching ratio was a factor three too
small:
\be
BR = \left\{
\begin{array}{lr} (1.7\pm 0.3) \times 10^{-6} &(\mbox{NA31}) \\
(1.86 \pm 0.60 \pm 0.60)\times 10^{-6}        &(\mbox{E731}) \\
0.67 \times 10^{-6} &~~~~~~O(p^4) \; \; ,
\end{array} \right.
\ee therefore requiring large $O(p^6)$ corrections.  The calculations at
order $p^6$\cite{Kpigg_p6} have considered only the (possibly dominant)
corrections to the pion loops, and added to this a polynomial contribution:
\begin{eqnarray*}
A&=&{4 \over s} (s-M_\pi^2) \tilde F\left({s \over M_\pi^2} \right) +
4 a_V\frac{3M_K^2-s-M_\pi^2}{M_K^2} + \ldots \\
B&=&\tilde G\left({s \over M_\pi^2} \right) -8 a_V + \ldots \; \; ,
\end{eqnarray*}
where $\tilde F(x)$ and $\tilde G(x)$ also come from $\pi \pi$ intermediate
state in the $s$ channel. To get into agreement with the experiment one
needed to have a large and negative $a_V$: $BR=0.83 \times 10^{-6}$ with
$a_V=0$ and $BR=1.60 \times 10^{-6}$ with $a_V=-0.9$.  Also for the
spectrum, unitarity corrections alone were not sufficient (and actually
worsened the agreement), while an improved agreement with the data is
obtained only with $a_V\sim-0.9$.

The outcome of this $O(p^6)$ analysis is therefore a clear need for a very
large contribution from the polynomial part. Is this reasonable or does it
signal a serious failure of the chiral expansion in this case?  Another way
to formulate this question is to ask whether we understand the dynamical
reason to have such a large constant. It is important to give a historical
perspective here. In '93 Cohen, Ecker and Pich in Ref.\cite{Kpigg_p4}
described the situation as follows: ``Several model estimates of $a_V$ have
been made in the literature.  A fair summary of those attempt is that we
know neither the sign nor the magnitude of $a_V$.''  More recently
D'Ambrosio and Portol\'es\cite{DAP97} have built a Vector Resonance
Model that does indeed get the right sign and size for this constant:
$a_V^{DP} \simeq -0.72$. This number is now in amazing agreement with the
one extracted from a fit to the most recent data\cite{KTeV}.

Although D'Ambrosio and Portol\'es estimate of $a_V$ was only a
postdiction, it is reassuring to have an understanding of the size of this
constant. In fact it is not the only case where one can find this relative
size between the various contributions in the chiral expansion. A
well--known analogous example in the strong sector is the vector form
factor.  Its Taylor expansion around $s=0$ is usually defined as $ F_V(s) =
1 + 1/6 \langle r^2 \rangle^\pi_V s + c_V^\pi s^2 +O(s^3)$.  $c_V^\pi$
vanishes at order $p^2$, and can be predicted with no parameters at order
$p^4$. Nowadays it is known up to order $p^6$ \cite{BCT}: \bea c_V^\pi&=&
\frac{1}{960 \pi^2M_\pi^2 F_\pi^2} + \frac{1}{(16 \pi^2 F_\pi^2)^2} \left[
\mbox{``}\ln {M_\pi^2 \over \mu^2}\mbox{''}+ \mbox{``}l_i^r\mbox{''}
\right] + \frac{r^r_{V2}}{F_\pi^4} \\ &=&(0.62 + 1.96 + 1.3\cdot 10^{-4}
r^r_{V2}(M_\rho)) \gev^{-4} = 5.4 \gev^{-4}\;, \nonumber \eea where the
latter value is determined experimentally. Here also: \newline
\noindent i) the order $p^4$ parameter--free prediction fails badly;
\newline \noindent ii) there are large $O(p^6)$ unitarity corrections;
\newline \noindent iii) but even larger $O(p^6)$ polynomial contributions,
coming from the $\rho$ resonance.

\subsection{$K \to \pi l^+ l^-$}

Already in 1987 Ecker, Pich and de Rafael\cite{EPdR} calculated the
amplitude of this decay mode at order $p^4$. The amplitude depends on one
(unknown) low--energy constant. Since we have two leptonic modes we can fix
the constant in one of the two and then predict the other.  Years after the
theoretical prediction data have appeared for both leptonic modes: BNL-E777
on the electron mode\cite{BNL-E777} and BNL-E787 on the muon mode
\cite{BNL-E787}. There are various interesting aspects in this decay mode,
and I refer the interested reader to a recent paper where one can also find
reference to the relevant literature\cite{DAEIP}. Here I only want to
discuss one particular number, the ratio of the width in the two modes. The
experimental measurements gave  
\be
R={\Gamma(K^+ \to \pi^+ \mu^+ \mu^-) \over \Gamma(K^+ \to \pi^+ \mu^+
\mu^-) } =0.167 \pm 0.036 
\ee 
which is $2 \sigma$ away from the CHPT value
$(\sim 0.24)$.  Again an example of a quantity subject to large corrections
from higher orders? D'Ambrosio, Ecker, Isidori and Portol\'es\cite{DAEIP}
have extended the $O(p^4)$ analysis to include the main $O(p^6)$ effects:
unitarity corrections (reliably calculable) and polynomial contributions
(estimated with theoretical modelling).  Their conclusion is that there is
no room for large corrections: 
$ R= \simeq 0.23$. A recent new measurement\cite{muonmode} of the muon mode
has brought the experimental number into agreement with the theoretical
prediction. 

\section{Conclusions}
Chiral perturbation theory is an essential tool to describe kaon physics,
an extremely interesting physics field that continues to have a very deep
impact on our knowledge and understanding of the physics of the Standard
Model, and also on what lies beyond it. DA$\Phi$NE and KLOE, as well as
fixed target experiments\cite{lowe,kettell}, are now starting to explore
this field at a very high precision level. I have quickly reviewed the
current status of this effective field theory, with special emphasis on a
few physics issues that are relevant for DA$\Phi$NE.

In some of the examples I have discussed theory is ahead of experiment, and
provides a solid and accurate prediction: the forthcoming experiments will
thoroughly test the theory and in particular (in $\pi \pi$ scattering) a
very fundamental aspect of the strong interactions, the structure of the
chiral symmetry breaking. In other cases experiment is ahead of theory, and
is providing essential informations for our understanding of the strong and
weak interactions, and their interplay.

\section*{Acknowledgements}
I warmly thank the organisers for the invitation to such an interesting and
successful conference.


\begin{thebibliography}{99}

\bibitem{handbook}
L.~Maiani, G.~Pancheri and N.~Paver (eds.), The Second DA$\Phi$NE Physics
Handbook (INFN, Frascati, 1995).

\bibitem{weinberg79}
S. Weinberg, Physica {\bf A 96}, 327 (1979).

\bibitem{GL} 
J. Gasser and H. Leutwyler, Ann. Phys. (N.Y.) {\bf 158}, 142
(1984), and Nucl. Phys. {\bf B 250}, 465 (1985).

\bibitem{BGS}
S. Bellucci, J. Gasser and M. Sainio, Nucl. Phys. {\bf B 423}, 80 (1994); 
{\bf B 431},413 (1994) (E).

\bibitem{ABT}
G.~Amoros, J.~Bijnens and P.~Talavera, \hepph{9912398}.

\bibitem{lagrp6}
J.~Bijnens, G.~Colangelo and G.~Ecker, JHEP {\bf 9902} 020 (1999),
\hepph{9907333}, Ann. Phys. (N.Y.), in press.

\bibitem{urech}
R.~Urech Nucl. Phys. {\bf B 433}, 234 (1995);
H.~Neufeld and H.~Rupertsberger Z. Phys. {\bf C71}, 131 (1996).

\bibitem{knecht}
M.~Knecht, H.~Neufeld, H.~Rupertsberger, and P.~Talavera, \hepph{9909284}.

\bibitem{lowe}
J.~Lowe, these proceedings.

\bibitem{kettell}
; S.~Kettell, these proceedings.

\bibitem{omega}
G.~Ecker, G.~M\"uller, H.~Neufeld, and A.~Pich, \hepph{9912264}

\bibitem{reviews}
J.~Bijnens Int. J. Mod. Phys. {\bf A8}, 3045 (1993); J.F. Donoghue,
E. Golowich, B.R. Holstein, ``Dynamics of the Standard Model'' 
Cambridge Univ. Pr. (1992);
G.~Ecker, Prog. Part. Nucl. Phys. {\bf 35}, 1 (1995);
J.~Gasser \hepph{9912548};
H.~Leutwyler, \hepph{9406283};
A.~Pich \hepph{9806303};
E. de Rafael, Boulder TASI 0015 (1994), \hepph{9502254}.

\bibitem{BCEGS}
J.~Bijnens, G.~Colangelo, G.~Ecker, J.~Gasser and M.~Sainio,
Phys. Lett. {\bf B 374}, 210 (1996), Nucl. Phys. {\bf B 508}, 263 (1997). 

\bibitem{weinberg66}
S.~Weinberg, Phys. Rev. Lett. {\bf 17}, 616 (1966).

\bibitem{GL83}
J.~Gasser and H.~Leutwyler, Phys. Lett. {\bf B 125}, 325 (1983).

\bibitem{GC95}
G.~Colangelo, Phys. Lett. {\bf B 350}, 85 (1995); {\bf B 361}, 234 (1995) (E).

\bibitem{rosselet}
L.~Rosselet et al. Phys. Rev. {\bf D 15}, 574 (1977).

\bibitem{BFP}
J.L.~Basdevant, C.D.~Froggatt, and J.L.~Petersen, Nucl. Phys. {\bf B 72},
413 (1974).

\bibitem{Roysolution}
B.~Ananthanarayan, G.~Colangelo, J.~Gasser, H.~Leutwyler and G.~Wanders,
work in progress.

\bibitem{hoogland}
W.~Hoogland et al. Nucl. Phys. {\bf B 126}, 109 (1977).

\bibitem{CGL}
G.~Colangelo, J.~Gasser, and H.~Leutwyler work in progress.

\bibitem{KMSF}
M.~Knecht, B.~Moussallam, J.~Stern and N.H.~Fuchs, Nucl. Phys. {\bf B 457}, 
513 (1995).

\bibitem{Deser}
S.~Deser, M.L.~Goldberger, K.~Baumann, and W.~Thirring, Phys. Rev. {\bf
  96}, 774 (1954).

\bibitem{Hadatom99}
For a summary of the present situation and reference to the relevant
literature, see
J.~Gasser, A.~Rusetsky, and J.~Schacher, \hepph{9911339}.

\bibitem{bern_atom}
A.~Gall, J.~Gasser, V.E.~Lyubovitskij, and A.~Rusetsky, Phys. Lett. {\bf B
  462}, 335 (1999);
J.~Gasser, V.E.~Lyubovitskij, and A.~Rusetsky, \hepph{9910438}.

\bibitem{CL}
W.E.~Caswell and G.P.~Lepage, Phys. Lett. {\bf B 167}, 437 (1986).

\bibitem{weak_lagr}
J.~Kambor, J.~Missimer, and D.~Wyler, Nucl. Phys. {\bf B 346}, 17 (1990);
G.~Ecker, J.~Kambor, and D.~Wyler, Nucl. Phys. {\bf B 394}, 101 (1993). 

\bibitem{Kpigg_p4}
G.~Ecker, A.~Pich, and E.~de Rafael, Phys. Lett. {\bf B 189}, 363 (1987);
G.~Cappiello, and G.~D'Ambrosio, Nuovo Cim. {\bf 99A}, 155 (1988).

\bibitem{Kpigg_p6}
G.~Cappiello, and G.~D'Ambrosio, and M.~Miragliuolo Phys. Lett. {\bf B 298},
423 (1993);
A.~Cohen, G.~Ecker, and A.~Pich, Phys. Lett. {\bf B 304}, 347 (1993);
J.~Kambor, and B.R.~Holstein, Phys. Rev. {\bf D 49}, 2346 (1994).

\bibitem{E731}
V.~Papadimitriou et al. (E731) Phys. Rev. {\bf D44}, 573 (1991).

\bibitem{NA31}
G.D.~Barr et al. (NA31) Phys. Lett. {\bf B 284}, 440 (1992). 

\bibitem{DAP97}
G.~D'Ambrosio, and J.~Portol\'es, Nucl. Phys. {\bf B 492}, 417 (1997).

\bibitem{KTeV}
A.Alavi-Arati et al. (KTeV), Phys. Rev. Lett. {\bf 83}, 917 (1999).

\bibitem{BCT}
J.~Bijnens, G.~Colangelo and P.~Talavera, JHEP {\bf 9805}, 014 (1998).

\bibitem{EPdR}
G.~Ecker, A.~Pich, and E.~de Rafael, Nucl. Phys. {\bf B 291}, 692 (1987).

\bibitem{BNL-E777}
C.~Alliegro et al. (E777) Phys. Rev. Lett. {\bf 68}, 278 (1992).

\bibitem{BNL-E787}
S.~Adler et al. (E787) Phys. Rev. Lett. {\bf 79} 4756 (1997).

\bibitem{DAEIP}
G.~D'Ambrosio, G.~Ecker, G.~Isidori, J.~Portol\'es JHEP {\bf 9808}, 004
(1998). 

\bibitem{muonmode}
H.~Ma et al. (E865) \hepex{9910047}

\end{thebibliography}
\end{document}